\documentclass[fleqn,usenatbib]{mnras}

\usepackage{newtxtext,newtxmath}

\usepackage[T1]{fontenc}

\DeclareRobustCommand{\VAN}[3]{#2}
\let\VANthebibliography\thebibliography
\def\thebibliography{\DeclareRobustCommand{\VAN}[3]{##3}\VANthebibliography}

\newcommand{\Msun}{\mbox{M$_{\odot}$}}

\newcommand{\lppr}{\stackrel{<}{\scriptstyle \sim}}
\newcommand{\lappr}{\raisebox{-0.4ex}{$\lppr$}}

\usepackage[usenames,dvipsnames]{xcolor}
\usepackage{color,colortbl}
\usepackage{soul}

\usepackage{graphicx}	
\usepackage{amsmath}	

\title[Most EL\,CVn binaries are triples]{Most EL\,CVn systems are inner binaries of hierarchical triples}
\author[F. Lagos et al.]{
F. Lagos$^{1,2,3}$\thanks{E-mail: felipe.lagos@postgrado.uv.cl},
M.R. Schreiber$^{2,4}$,
S.G. Parsons$^{5}$,
B.T. G\"ansicke$^{6}$,
N. Godoy$^{1,2}$
\\
$^1$Instituto de F{\'i}sica y Astronom{\'i}a, Universidad de Valpara{\'i}so, Valpara{\'i}so, Chile\\
$^2$Millennium Nucleus for Planet Formation, NPF, Valpara{\'i}so, Chile\\
$^3$European Southern Observatory (ESO), Alonso de Cordova 3107, Vitacura, Santiago, Chile\\
$^4$Departamento de F\'isica, Universidad T\'ecnica Federico Santa Mar\'ia, Av. Espa\~{n}a 1680, Valpara\'iso, Chile \\
$^5$Department of Physics and Astronomy, University of Sheffield, Sheffield S3 7RH, UK\\
$^6$Department of Physics, University of Warwick, Coventry, CV4 7AL, UK
}

\date{Accepted XXX. Received YYY; in original form ZZZ}

\pubyear{2020}

\begin{document}
\label{firstpage}
\pagerange{\pageref{firstpage}--\pageref{lastpage}}
\maketitle

\begin{abstract}
In spite of their importance for modern astronomy, 
we do not fully understand how close binary stars containing at least one white dwarf form from main sequence binary stars. 
The discovery of EL\,CVn binaries, close pre-white dwarfs with A/F main sequence star companions, offers now the unique possibility to test models of close compact binary star formation. Binary evolution theories predict that these EL\,CVn stars descend from very close main sequence binaries with orbital periods shorter than 3\,days. 
If this is correct, nearly all EL\,CVn stars
should be inner binaries of hierarchical triples because more than 95 per cent 
of very close main sequence binaries (the alleged progenitor systems) are found to be hierarchical triples.
We here present SPHERE/IRDIS observations of five EL\,CVn binaries, finding in all of them tertiary objects, as predicted.
We conclude that EL\,CVn systems are inner binaries of hierarchical triples and indeed descend from very close main sequence binaries that experience stable mass transfer.  
\end{abstract}




\section{Introduction}

A large variety of close white dwarf binaries exist, including objects as interesting 
as super soft X-ray sources, catalcysmic variables (CVs) and their detached progenitors
\citep[e.g.][]{schreiber+gaensicke03-1} or double white dwarf binaries \citep{han98-1}. 
For all these close white dwarf binaries, theories for their formation and evolution, in most cases involving common envelope evolution \citep{webbink84-1,zorotovicetal10-1}, have been developed. 

However, as our prescriptions for mass transfer interactions and angular momentum loss are relatively simple conservation equations containing a number of neither theoretically nor observationally well constrained parameters, binary population simulations are unable to reliably predict detailed characteristics of most white dwarf binary populations. In fact, in several cases, the predictions of theoretical binary population models strongly disagree with the observations. 
To provide just three examples, current models are unable to produce double degenerate systems via a combination of two common envelope phases \citep[][]{nelemans+tout05-1} and require either stable mass transfer or additional energy sources contributing during 
common envelope evolution \citep{webbink08-1}. No model for supernovae Ia, the thermonuclear explosion of a white dwarf growing in mass through binary star interactions, is able to reproduce the observed delay time distribution 
\citep[e.g.][]{yungelson+kuranov17-1}. Even worse, in CVs several predictions of the standard model for their evolution drastically disagree with the observations \citep[e.g.][]{zorotovicetal11-1}. While some of these problems are solved in a recently suggested revision of the model \citep{schreiberetal16-1}, others remain \citep{palaetal17-1,bellonietal20-1}. 

To test and eventually calibrate theoretical models of close white dwarf binary formation it would be ideal to 
have reliable and concrete information about the main sequence binary progenitors for a given type of white dwarf binary. 
In general, this is illusory. However, here we present, to the best of our knowledge, the first case where such information is indeed available: EL\,CVn binaries. 

EL\,CVn-type binaries are eclipsing binaries that contain an A- or F-type star and a very low mass (typically 0.2 solar masses) helium white dwarf precursor (pre-Helium WD). The orbital periods of EL\,CVn stars are typically very short, i.e. 1-3.\,days. A number of these compact binary stars have been discovered in the Wide Angle Search for Planets, Kepler photometric surveys, 
or with the Palomar Transient Factory \citep{maxtedetal14-1, vanroesteletal18-1}, and have been found to show different types of pulsations
\cite[e.g.][]{maxtedetal14-2}. 
 
What makes EL\,CVn binaries special for close white dwarf binary evolution theories is the following.  
While most white dwarfs and pre-white dwarfs in close binaries are assumed to have formed through common envelope evolution, according to current theories, EL\,CVn stars must form from dynamically stable mass transfer when the more massive star of the initial main sequence binary was at the end of its main sequence lifetime or just entered the sub-giant branch \citep{chenetal17-1}. This condition is required as otherwise the core of the more massive star will grow to masses exceeding measurements of pre-white dwarfs in EL\,CVn binaries.
Such an early start of mass transfer implies that the orbital period of 
the progenitor binary star system must have been shorter than $\sim$\,3 days, as the radius of the primary star was still close to that of a main sequence star when mass transfer began. For longer orbital periods, 
the sub-giant star is simply too small to fill its Roche-lobe. 

The prediction of very short orbital periods for the progenitor main sequence binary stars has fascinating consequences. 
Virtually all close main sequence stars with orbital periods below $\sim\,3$ days are known to be the inner binaries of hierarchical triple systems \citep{tokovininetal06-1}. Even if the mass transfer of the inner binary was not conservative and a certain amount of mass was expelled from the inner binary (thereby increasing the orbit of the tertiary), the third companions should still be there after the mass transfer phase. This implies that, if current theories for the formation of EL\,CVn stars are correct, virtually all EL\,CVn stars must be the inner binaries of hierarchical triples. 

We here present SPHERE \citep{Beuzit_2019} observations of five EL\,CVn binaries and indeed find strong candidates for tertiary objects in all of them. 
We conclude that EL\,CVn binaries form from the inner and very close binaries 
of hierarchical triple systems in perfect agreement with the formation channels described in \citet{chenetal17-1}.

\section{SPHERE Observations and data reduction}

The separations of the potential tertiary companions to the inner (EL\,CVn) binary are expected to be in the rage from 10 to 10000\,au \citep{tokovininetal06-1}. With the spatial resolution of roughly 100 mas and the field of view of SPHERE/IRDIS (11 x 11 arcsec$^2$, \citealt{2008SPIE.7014E..3LD}), 
our targets must be at distances between $\approx$200$-$1800\,pc in order to cover a large range of the predicted separations for all our targets. We selected five EL\,CVn binaries from the sample of \citet{maxtedetal14-1} with distances between 420$-$1600\,pc 
and magnitudes in the 2MASS $H$ 
band between 9.7 and 11.9. 
We used \textit{Gaia} distances except for TYC\,6736. 
For the latter, the \textit{Gaia} measurement implies an unrealistically large distance (2080 pc). In fact, the poorly constrained paralax (0.46$\pm$0.23 mas) derived by \textit{Gaia} may already indicate the presence of a 
companion. Instead of the \textit{Gaia} value, 
we used the distance from RAVE data release 5 \citep{RAVE-DR5} for TYC\,6736. Table\,\ref{tab_1} summarizes the
parameters of our targets. Thanks to the extreme contrast of 8$-$14\,mags that can be reached with SPHERE our observations are even sensitive to low-mass ($\approx 0.1 \Msun$) stellar companions to our most distant objects. 
\begin{table}
	\caption{Distances, 2MASS \textit{H} photometry, orbital period, mass ratios and proper motion of all EL\,CVn stars studied in this work. Orbital periods and mass ratios are obtained from \protect\citet{maxtedetal14-1}. Distances are obtained from \protect\cite{bj-gaia}, except for TYC\,6736, whose distance has been obtained from data release\,5 of the RAVE survey \protect\citep{RAVE-DR5}.}
	\label{tab_1}
	\resizebox{\columnwidth}{!}{\begin{tabular}{llcccc} 
		\hline
	Target & Distance [pc]  & $H\mathrm{_{2MASS}}$ & $\mathrm{P_{orb}}$&$q$     & Proper Mot.  \\
	       &\hspace{0.5cm}$\pm$&             $\pm$    &       [d]             & $\pm$  &      [mas/yr] \\
	 \\ 
	\hline
	TYC 5204-1575-1&745.37  &11.308             &1.29 & 0.130    &12.8 \\
	               &    46    &0.022              &     & 0.026    & \\
	
    TYC 9337-2511-1&1572.55 &11.874  &1.162&0.148  &9.5  \\
    	           &     66   &0.023   &     & 0.026 & \\
    
    TYC 6736-69-1  &424     & 9.924 &2.173&0.136 &13.7 \\
        	       &85        &0.021  &     & 0.041& \\
    
    TYC 6631-538-1 &540.07  &10.432 &0.901&0.143  &8.9 \\
            	   &13        &0.027  &     & 0.027& \\
    
    TYC 5450-1192-1&447.37  & 9.748  &0.793&0.176  &12.8 \\
                	        & 8.5       &0.022&     & 0.012& \\

       \hline
	\end{tabular}}
\end{table}

The five objects that we selected were observed with the high contrast imager SPHERE between 2019 April 9 and August 13. Acquisition of direct imaging was made with IRDIS in the dual band imaging mode \citep{2010lyot.confE..48V} using the broad band $H$ filter ($\lambda_\mathrm{H}=1625$ $\pm$ 290 $\mu$m). Furthermore, the pupil tracking mode was implemented in order to perform angular differential imaging \citep[ADI,][]{2006ApJ...641..556M}. 
We used the N-ALC-YJH-S coronagraph.

\begin{figure*}
\centering
  \includegraphics[width=0.87\columnwidth]{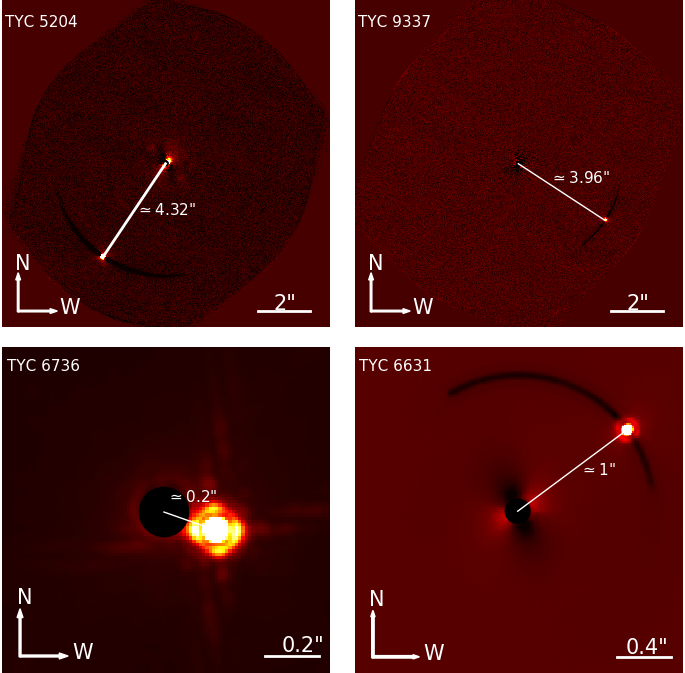}
 \hfill
  \includegraphics[width=0.855\columnwidth]{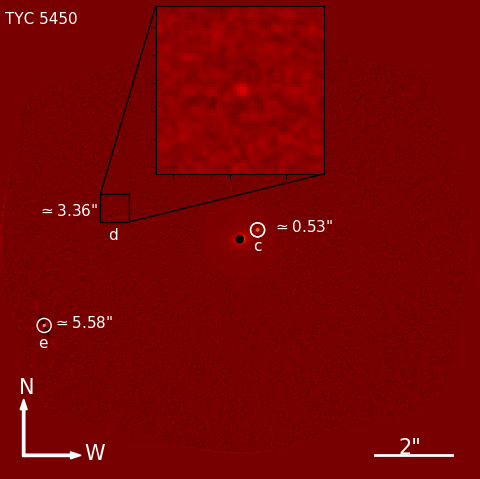}
\caption{SPHERE/IRDIS images of the five EL\,CVn stars that we observed. 
In all cases we find at least one potential tertiary. The magnitude difference obtained for the possible companions extend from  1.3 to 14\,mags. 
The probabilities for being background objects are small in all cases with values in the range of $0.0002-3$\% (except for TYC\,5450 d and e). 
}
\label{fig:irdis}
\end{figure*}

At the beginning and end of each observing sequence, we obtained \textit{flux} (with the neutral density filter ND\,1
to avoid saturation) 
and \textit{star centre} calibration images, adding one more exposure for the latter in the middle of the sequence. 
A total of 
112 science images were taken with an exposure time of 32\,s for each target except 
TYC\,6736 for which only 66 science images were taken due to bad weather. 
The IRDIS data were first pre-processed (sky background subtraction,
flat-fielding, bad-pixels correction) with the \textsc{vlt/sphere} python package\footnote{https://github.com/avigan/SPHERE}. The frames were recentred based on star centre exposures using the four satellite spots. 
After pre-processing, and without any post-processing technique to remove speckle patterns produced by the coronograph, we detected at least one potential companion in all five EL\,CVn binaries. We used the principal component analysis (PCA) algorithm available in the \textsc{Vortex Image Processing} \citep[\textsc{vip},][]{Gomez_Gonzalez_2017} python package to look for fainter companions, only finding one extra candidate in TYC\,5450 (TYC\,5450 d, see Fig. \ref{fig:irdis}).

Making use of the strong signal of our detections, we derotated and median-collapsed the science images to perform aperture photometry and obtained their relative magnitudes $\Delta H_{\mathrm{mag}}$ with respect to the central binary. Since the object detected in TYC\,6736 surpassed the IRDIS saturation threshold, flux calibration images were used to obtain $\Delta H_{\mathrm{mag}}$. As a complementary approach we also used the negative fake companion technique coupled with a Markov chain Monte Carlo algorithm available in \textsc{vip} to calculate $\Delta H_{\mathrm{mag}}$, without finding any discrepancy between both methods.

\section{The tertiaries to \texorpdfstring{EL\,CVn}{EL CVn} stars}

In all five cases, our SPHERE observations revealed 
potential tertiary companions (Fig. \ref{fig:irdis}). 
In what follows we show that the probability for these detections to be background objects is very small
and estimate the possible nature of the companions.

\subsection{Excluding background contamination}

We calculated the probability P($\Theta,m$) 
for each detection to be a chance alignment with a background source within an angular distance $\Theta$ following \citet{2000AJ....120..950B} as 
\begin{equation}
  P(\Theta,m_\mathrm{lim})= 1-e^{-\pi \Theta^2 \rho(m_\mathrm{lim})}, 
\end{equation}  
where $\rho(m)$ is the cumulative surface density of background sources down to a limiting magnitude m$_\mathrm{lim}$ (i.e. the magnitude of the detections). In order to calculate $\rho(m_\mathrm{lim})$, we used the Besan\c{c}on galaxy model
(\citealt{refId0}) to generate a synthetic 2MASS \textit{H} photometric catalogue of point sources within 1 square degree, centred on the coordinates of each target. To calculate $\mathrm{m_{lim}}$, we assumed that both SPHERE and 2MASS $H$ filters are identical, so that the relation between the SPHERE $\Delta H_{\mathrm{mag}}$ of our candidates and their apparent 2MASS $H$ magnitudes ($H\mathrm{_{2MASS}}$) is given by
\begin{equation}
  H_{\mathrm{2MASS}}=H_{\mathrm{arch,2MASS}} + 2.5\log(1+\alpha),
\end{equation} 
where $H_{\mathrm{arch,2MASS}}$ is the 2MASS archival magnitude of the EL\,CVn and $\alpha=10^{0.4\Delta H_{\mathrm{mag}}}$ the value of $\Delta H_{\mathrm{mag}}$ expressed in terms of the counts ratio between the EL CVn and the candidate. 
To avoid underestimating the value 
of P$(\Theta,m_\mathrm{lim})$, we added 0.5 magnitudes to $m_\mathrm{lim}$.
We note that discrepancies between 2MASS and SPHERE $H$ bands are clearly irrelevant. We used VOSA's \citep{2008A&A...492..277B} synthetic photometry tool coupled with the BT-NextGen spectral library \citep{Allard12}, finding differences of 0.06 magnitudes at most. 
Table \ref{tab_prob} lists the final apparent magnitudes and probabilities for being a background object for each candidate. %
\begin{table}
	\caption{Measured separations, magnitude differences, apparent magnitudes, projected separations, probabilities for chance alignment with a background source, mass and effective temperature for the seven detections.  
	}
	\label{tab_prob}
	\resizebox{\columnwidth}{!}{\begin{tabular}{llcccccc} 
		\hline
	Object & $\Theta$  & $\Delta H_\mathrm{_{mag}}$& Proj. Sep. [au] & $P(\Theta,m_\mathrm{lim})$ & Mass [\Msun] \\
	       & [arcsec]& $H_{\mathrm{2MASS}}$& $\pm$& &T$_\mathrm{eff}$ [K] \\
	 
	\hline
	TYC\,5204 c& 4.32 & 4.988 $\pm$ .018 & 3193 & 0.023& 0.3-0.4 \\
	           &      & 16.29 $\pm$ .08  &  198     &      &3400-3500 \\
    TYC\,9337 c& 3.96 & 6.946 $\pm$ .016 & 6227 & 0.031& 0.2-0.3\\
               &      & 18.82 $\pm$ .08  &  261      &      & 3200-3400\\
    TYC\,6736 c& 0.2  & 1.330 $\pm$ .030 & 88  & $2 \cdot 10^{-6}$&$\approx 0.9$\\
            &      & 11.53 $\pm$ .09  &    18    &      & 5300-5500\\
    TYC\,6631 c& 1.0  & 3.595 $\pm$ .008 & 544 & $4\cdot 10^{-4}$&$\approx 0.6$\\
               &      & 14.03 $\pm$ .07  &    13    &      & 3700-4000\\
    TYC\,5450 c & 0.53 & 4.701 $\pm$ .014& 235 & 1.6 $\cdot 10^{-4}$&$\approx 0.5$\\
                &      & 14.45 $\pm$ .07  &    5    &      &$\approx 3700$ \\
                
	TYC\,5450 d & 3.36 & 13.86 $\pm$ .15 & 1503 & 0.089& $<0.07$ \\ 
	            &      & 23.6 $\pm$ .2   &   28     &      & $<1600$ \\
	          
	TYC\,5450 e & 5.58 & 10.402 $\pm$ .120 & 2497 & 0.15&$<0.07$ \\ 
	            &      &20.15 $\pm$ .18   &    47    &      & $<1600$ \\	
       \hline
	\end{tabular}}
\end{table}
Combining the obtained probabilities we derive a probability for all five EL\,CVn binaries to be triple 
systems or higher multiples of $0.95$, i.e. with 2$\sigma$ significance we measured that five out of five EL\,CVn stars are triples or higher multiples. 

For any assumed true triple fraction of EL\,CVn stars we can now 
calculate the significance for rejecting this hypothesis. Given an assumed triple fraction $p$, the probability of measuring five out of five is simply $p^5$. This implies that we can say with one (two) sigma significance that the true fraction of EL\,CVn stars exceeds 80 (55) per cent.

In Fig. \ref{fig:fontok} we compare our result with the triple fraction of sun-like stars as a function of orbital period. 
Apparently, the measured high fraction of tertiaries in EL\,CVn binaries corresponds to those of close 
binary stars with periods below $3$\,days, in perfect agreement with theoretical predictions. 

In general one could further strengthen this result by 
confirming common proper motions of the candidates with 
respect to the central EL\,CVn binary. 
According to the proper motions listed in 
Table\,\ref{tab_prob} and the IRDIS pixel scale (12.25 mas), 
the time between observations would need to be at least 3 yr. 
However, such second epoch observations would only slightly 
improve an already highly significant result given that 
for two objects independent evidence for the companion nature of our candidates is already provided by \textit{Gaia}. 
As mentioned previously, the large uncertainty and unrealistic value of the \textit{Gaia} paralax of TYC\,6736 is 
consistent with the presence of a companion. In the case of TYC\,5204, \textit{Gaia} even detected the companion and, despite relatively large uncertainties, the \textit{Gaia} astrometric solution is consistent with common proper motion.
Given this independent evidence, and the very low background probabilities, our result would remain nearly unchanged 
even if the Galaxy model we used was significantly underestimating the background object probability.
%

\subsection{Potential nature of the companions}

The optical and IR emission of EL\,CVn binaries is dominated by 
their A (or F) star. If our detections are physically bound to the EL CVn binary, we would expect they were later G, K or M type dwarfs. Although a full characterization of the potential companions is impossible given the limited data currently available, we can derive a rough estimate of their masses and temperatures by using isochrones for main sequence low-mass stars \citep{2015A&A...577A..42B}.

\citet{chenetal17-1} showed that EL CVn binaries with orbital periods shorter than 2.2 days descend from main sequence binaries where the most likely mass for the pre-Helium white dwarf progenitor is in the range $1.15-1.2$\Msun. As the mass transfer that led to the formation of the EL\,CVn system started after/at the end of main sequence evolution, the age of our systems should be at least $\approx 5-7$ Gyr. By comparing the derived distance-modulus corrected $H_{\mathrm{2MASS}}$ magnitude with those values in the isochrones between 5 and 10 Gyr, the companions for TYC\,5204, TYC\,9337, TYC\,5450 and TYC\,6631 are consistent with with M and K type stars, while for TYC\,6736 the closest companion is likely a G type star (Table\,\ref{tab_prob}).

\section{Discussion} 

Binary evolution theory predicts that EL\,CVn stars descend from very close main sequence binary stars. Multiplicity surveys show that practically all such close main sequence binaries host a distant tertiary, i.e. they are the inner binaries of triple systems.  
Our observations with SPHERE/IRDIS of five EL\,CVn confirm this prediction: we find strong companion candidates in all five systems. 
In what follows we discuss the implications of our finding on the evolutionary history and the future of EL\,CVn stars.  

\subsection{The evolutionary history of \texorpdfstring{EL\,CVn}{EL CVn} triples} 

EL\,CVn binaries descend from hierarchical main sequence triple star systems where the inner binary has a very short orbital period of $\lappr\,3$\,days. The formation of such close binaries has been a mystery for 
decades as they cannot form directly via fragmentation of molecular clouds or protostellar discs. \citet{fabrycky_and_tremaine_2007} and \citet{naoz+fabrycky14-1} proposed that Kozai-Lidov oscillations coupled with tidal friction could be responsible for the formation of these close binaries. 
However, \citet{moe+kratter18-1} showed that 
the close binary fractions of very young stars is consistent with those of 
field stars and that Kozai-Lidov oscillations are too slow to reproduce the observations. 
Instead, they suggest that energy dissipation due to interactions with the primordial disc  
during the pre-main sequence is the main mechanism driving the formation of inner binaries with periods less than 10 days. 

\begin{figure}
\begin{center}
 \includegraphics[width=0.99\columnwidth]{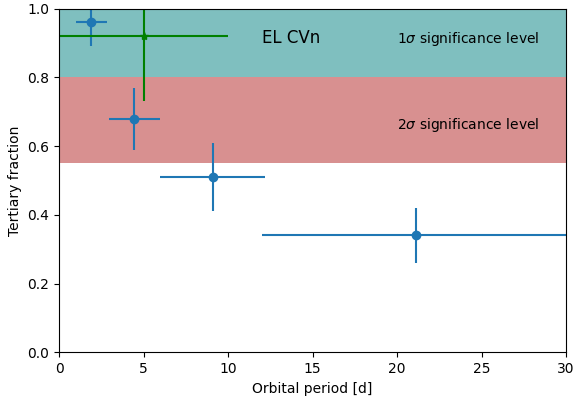}
 \caption{Triple fraction for stellar binaries (blue) and hot Jupiter/Brown dwarfs (green) from \citet{tokovininetal06-1} and \citet{fontaniveetal19-1}. 
 In both cases the fraction of systems with tertiary companions is very large, suggesting that close binaries (including those with massive Jupiter or brown dwarf companion) form through disc fragmentation \citep[see also][]{moe+kratter19-1}. The teal and red shaded regions represent the $1\sigma$ and $2\sigma$ significance level for rejecting the hypothesis that the tertiary fraction of EL\,CVn stars is less than 80 and 55 per cent respectively. The large triple fraction clearly implies that EL\,CVn stars descend mostly from close inner binaries of hierarchical triples. 
 } 
 \label{fig:fontok}
 \end{center}
\end{figure}

Key information concerning the origin of very close binary stars may come from combining 
statistics of stellar multiples with those of the most massive planets and brown dwarfs. 
Recently, \citet{fontaniveetal19-1} showed that the fraction of tertiaries in systems with 
close substellar companions is even higher than in the case of stellar binaries. We compare the 
triple fraction of EL\,CVn stars with both samples as a function of orbital period in Fig.\ref{fig:fontok}. 
As also noted by \citet{moe+kratter19-1}, the similar large fractions of systems with a
distant tertiary indicate a similar formation mechanism which is most likely disc fragmentation followed by disc migration. Why this disc fragmentation and subsequent migration occurs nearly only in hierarchical triples is not entirely clear. Either the total initial mass required for this to work is large enough that it is virtually always accompanied by core fragmentation, or 
a close companion formed through core fragmentation is triggering disc fragmentation at an early stage. In any case, we conclude that according to current theories of close binary formation, EL\,CVn systems are triple systems with a close inner binary formed through disk fragmentation and subsequent migration. 
After the main sequence triple has formed, it is assumed to evolve quietly until the more massive star of the inner binary 
moves off the main sequence and crosses the Hertzsprung gap. This star then fills its Roche-lobe before entering the first giant branch and thermal time scale mass transfer starts. It depends then on the thermal time scales and on the masses of both stars that to what degree this dynamically stable mass transfer remains conservative and how much the orbital period is increased during the mass transfer phase. Mass transfer stops when the envelope of the initially more massive star has been fully stripped off. The result is an inner binary consisting of the exposed low-mass core and a secondary that is more massive than it initially was 
\citep[][]{chenetal17-1}. The tertiary is affected little by the evolution of the inner binary. In most cases it simply spirals out depending on the total mass lost by the inner binary: an EL\,CVn triple system, as the ones we observed, is born. 

\subsection{The future of \texorpdfstring{EL\,CVn}{EL CVn} triples} 

The evolutionary phase the triple system appears as an EL\,CVn binary with a distant tertiary is probably relatively short. 
Given the short periods of the EL\,CVn binaries and in particular of the sample observed by us (see Table \ref{tab_1}), a second phase of mass transfer is unavoidable. 
For shorter periods the main sequence star either fills its Roche-lobe while still on the main sequence or in the Hertzsprung gap. In both cases it will
not have a deep convective envelope and mass transfer will be stable for mass 
ratios as large as 3-5 \citep[see][their Fig. 8 and 9]{geetal15-1}. Nevertheless, the mass ratios of the EL\,CVn stars we observed exceed this limit (table \ref{tab_1}, using $1/q$ for a proper comparison) and it is therefore likely that the second phase of mass transfer will be dynamically unstable and lead to common envelope evolution. As the orbital energy that is available will not be sufficient to expel the entire envelope of the main sequence star, both stars will merge, hydrogen burning will start around the 
core until the envelope will be expelled. In other words, the triple systems we currently observe will evolve into wide binary systems with one component being a white dwarf. If a significant part of the secondary is expelled during this merger process, the white dwarf may not significantly grow in mass and remain in the low-mass white dwarf regime. Interestingly, such a system has recently been observed \citep{vosetal18-1} and the authors concluded that it most likely descended from a triple, in perfect agreement with our prediction for the future of close EL\,CVn binaries. 

EL\,CVn stars with slightly smaller mass ratios than the systems 
we observed will start thermal time scale mass transfer and evolve into double degenerate stars consisting of two low-mass white dwarfs. These EL\,CVn stars might therefore be the progenitors of observed double white dwarfs
\citep{boursetal14-1,parsonsetal20-1}.
EL\,CVn stars with significantly longer periods have not been observed but predicted to exist (though in small numbers). These objects will start mass transfer 
when the main sequence star evolved into a giant star which will lead to common envelope evolution. Due to the larger initial orbital period, these systems 
may survive the common envelope phase and form very close double degenerates. For both scenarios, the amount of mass lost during mass transfer or common envelope evolution would be, in most cases, not enough to unbound the third object, even in an \textit{impulsive mass-loss} regime \citep{Veras_2011}, i.e. the distant tertiary will move to a larger separation but remains bound to the system. As EL\,CVn systems are likely to evolve into systems containing extremely low-mass (ELM) white dwarfs, the sample established by the ELM survey \citep[e.g.][]{Brown_2010} might contain a significant number of triple systems. 
The outlined future evolution of EL\,CVn stars and the origin 
of ELM white dwarfs can be tested by searching for these tertiaries. 
To predict the precise relative number of such systems, triple population synthesis models are required which we intend to present in an upcoming publication. 

\section{Conclusions}
The characteristic configuration of EL\,CVn binaries, close A-F type dwarfs with a very low-mass pre-white dwarf evolving towards higher effective temperatures at nearly constant luminosity, offers a unique way to test the latest white dwarf binary 
formation theories. The narrow parameter space predicted 
for the progenitor systems, i.e. orbital periods 
below $\sim3$ days and white dwarf progenitors with masses $\simeq 1.15-1.20\Msun$, implies that (if theories are correct) nearly all EL\,CVn binaries must be the inner binary stars of hierarchical triples because virtually all main sequence binary stars with periods shorter than 3\,days are known to be the inner binaries of such triples. 

We performed SPHERE/IRDIS observations of five EL\,CVn stars and  indeed found very strong companion candidates in all five systems. Our results represent a unique and independent confirmation of the predictions of formation theories for EL\,CVn stars. 
Discussing the future of EL\,CVn binaries we found that EL\,CVn stars and their tertiaries either evolve into wide binaries with a low-mass white dwarf component or into 
triples with inner binaries consisting of at least one extremely low-mass (ELM) white dwarf.
EL\,CVn triples therefore represent a link between hierarchical main sequence triples with very close inner binary stars and a sub-population of systems containing ELM white dwarfs.

\section*{Acknowledgements}
FL, MRS, and NG thank for support from the ICM Millennium Nucleus for Planet Formation, NPF. 
FL is supported by an ESO studentship and MRS by FONDECYT (1181404).
SGP acknowledges support from the STFC Ernest Rutherford Fellowship.
BTG was supported by the UK STFC grant ST/P000495.
NG acknowledges support from project CONICYT-PFCHA/Doctorado Nacional/2017 folio 21170650. 
The data presented in this work have been obtained through ESO program 0103.D-0346(A). 

\section*{Data availability}
The data underlying this letter will be shared on reasonable request to the corresponding author.







%


\bsp	
\label{lastpage}
\end{document}